\documentclass{article} % For LaTeX2e
\usepackage{arxiv, times}

\usepackage{hyperref}       % hyperlinks

\usepackage{url}            % simple URL typesetting
\usepackage[utf8]{inputenc} % allow utf-8 input
\usepackage[T1]{fontenc}    % use 8-bit T1 fonts
\usepackage{booktabs}       % professional-quality tables
\usepackage{amsfonts}       % blackboard math symbols
\usepackage{nicefrac}       % compact symbols for 1/2, etc.
\usepackage{microtype}      % microtypography

\usepackage{graphicx}
\usepackage{subfigure}
\usepackage{amsmath}
\usepackage{amssymb}
\usepackage[font=small]{caption}
\usepackage{mathtools}
\usepackage{dsfont}
\usepackage{float}
\usepackage{multirow}
\usepackage{tabularx}
\usepackage{bm}
\usepackage{arydshln}
\usepackage{pifont}
\usepackage{algorithm}
\usepackage{todonotes}
\usepackage{courier}
\usepackage{algpseudocode}
\usepackage{bbm }
\usepackage{wrapfig}

\usepackage[nottoc]{tocbibind}
\usepackage{minitoc}

\usepackage{enumitem}
\setlist[itemize]{leftmargin=5mm, nolistsep}

\usepackage{color}
\usepackage[linewidth=0.5pt]{mdframed}

\author{Linghan Zhong\textsuperscript{1}, Ryan Lindeborg\textsuperscript{1}, Jesse Zhang\textsuperscript{1}, Joseph J. Lim\textsuperscript{2} \textsuperscript{3}, Shao-Hua Sun\textsuperscript{4} \\
\textsuperscript{1}USC, \textsuperscript{2}KAIST, \textsuperscript{3}AI Advisor at Naver AI
Lab, \textsuperscript{4}National Taiwan University
}
\newcommand{\mytitle}{\title{
Hierarchical Neural Program Synthesis
}}

\newcommand{\sun}[1]{{\color{blue}{\small\bf\sf [Sun: #1]}}}
\newcommand{\sunnote}[1]{}
\newcommand{\han}[1]{{\color{red}{\small\bf\sf [Han: #1]}}}
\newcommand{\jesse}[1]{{\color{green}{\small\bf\sf [JZ: #1]}}}
\newcommand{\ryan}[1]{{\color{cyan}{\small\bf\sf [Ryan: #1]}}}
\newcommand{\Skip}[1]{}

\newcommand{\method}[1]{HNPS}

\newcommand{\ie}{\textit{i}.\textit{e}.,\ }
\newcommand{\eg}{\textit{e}.\textit{g}.,\ }

\newcommand{\myfig}[1]{Figure \ref{#1}}

\newcommand{\mysecref}[1]{Section \ref{#1}}

\newcommand{\dotieconcat}[2]{% auxiliary macro, don't use it directly
  \text{\raisebox{.8ex}{$\smallfrown$}}%
}

\makeatletter
\newcommand\dslfontsize{\@setfontsize\dslfontsize\@viipt\@viiipt}
\makeatother

\newcommand{\myparagraph}[1]{\noindent \textbf{#1.}}
\newcommand{\vspacesection}[1]{\vspace{-0.0cm}
\section{#1}
\vspace{-0.0cm}}
\newcommand{\vspacesubsection}[1]{\vspace{-0.0cm}
\subsection{#1}
\vspace{-0.0cm}}

\usepackage{color}
\usepackage{xcolor}
\definecolor{codegray}{rgb}{0.5,0.5,0.5}
\iclrfinalcopy % Uncomment for camera-ready version, but NOT for submission.
\begin{document}

\doparttoc % Tell to minitoc to generate a toc for the parts
\faketableofcontents % Run a fake tableofcontents command for the partocs

\mytitle
\maketitle

\begin{abstract}
Program synthesis aims to
automatically construct human-readable programs 
that satisfy given task specifications, such as input/output pairs or demonstrations.
Recent works have demonstrated encouraging results
in a variety of domains, such as 
string transformation, tensor manipulation, 
and describing behaviors of embodied agents.
Most existing program synthesis methods 
are designed to synthesize programs from scratch,
generating a program token by token, line by line.
This fundamentally prevents these methods from scaling up to synthesize programs that are longer or more complex.
In this work, we present a scalable program synthesis framework
that instead synthesizes a program by hierarchically composing  programs.
Specifically, 
we first learn a task embedding space 
and a program decoder that can decode a task embedding into a program.
Then, we train a high-level module to comprehend the task specification (\eg input/output pairs or demonstrations) from long programs 
and produce a sequence of task embeddings,
which are then decoded by the program decoder
and composed to yield the synthesized program.
We extensively evaluate our proposed framework in
a string transformation domain with input/output pairs.
The experimental results demonstrate that 
the proposed framework can synthesize programs 
that are significantly longer and more complex than 
the programs considered in prior program synthesis works. Website at \url{https://thoughtp0lice.github.io/hnps_web/}

\end{abstract}

\vspacesection{Introduction}
\label{sec:intro}

% \sun{program synthesis: goal}
Program synthesis
aims to automatically synthesize a program 
structured in a domain-specific language (DSL)
that satisfies given task specifications.
Recently, encouraging results have been achieved across a variety of domains
such as string and tensor manipulation,
computer commands, 
graphics programs,
and describing embodied agent behaviors~\citep{devlin2017robustfill, balog2016deepcoder, bunel2018leveraging, nsd, lin2018nl2bash, sun2018neural, trivedi2021learning}.
% However, these methods generally synthesize programs token-by-token, which can make scaling to longer programs difficult.
% However, these methods are hard to scale as they synthesize programs token-by-token, making it difficult to synthesize longer programs and receiving weak training supervision that weights each token equally.

% \{limitation of current methods}
Most existing program synthesis methods are designed to sequentially synthesize programs---generating program tokens one by one based on previously produced tokens until the synthesis process has finished. However, without any inductive bias on the program space, synthesizing desired programs token by token can become difficult when we scale from short programs with simple behaviors to longer programs with more complex behaviors. Furthermore, these methods often learn to maximize the likelihood of each token with losses that weigh their importance in the program equally regardless of how they can affect the program's behavior. % regardless of its importance in the program. 
Such a training scheme suffers from the \emph{program aliasing problem}---where different tokens lead to the same program behavior---and having \emph{weak supervision for critical tokens}---single tokens whose replacement will greatly change the program output. For example, while a program output may not change significantly when a non-critical token is replaced with a different one, such as "nullptr" and "NULL" which are mostly interchangeable in C++, the output is completely changed when replacing input values of key functions (\ie the position index token of an array is changed from a[0] to a[1]). 
% The program aliasing problem is characterized by a generated program that has correct behavior for the limited samples in the task specification but does not truly represent the desired behavior and is different from the ground truth program. 
Thus, we argue that this token-by-token paradigm for generation and training fundamentally prevents current program synthesis methods from scaling up to synthesizing more complex programs. 
% our goal: to synthesize long programs
% our method: two-stage framework

% \sun{related work}

% \sun{our method}
In this work, we aim to develop a program synthesis framework 
that can scale up to longer programs 
which can induce more complex behaviors. 
Our key idea is to compose shorter programs with simple behaviors
to form longer programs with more complex behaviors.
% To this end, we take advantage of the fact that long programs can be composed of shorter programs with simple behaviors. 
% Our proposed method, the Hierarchical Neural Program Synthesizer (HNPS),
% is designed to learn to synthesize programs 
% by composing sequences of short programs to form longer programs. 
% {why hnps hard}
However, the space of meaningful short programs is intractable in most cases -- searching for correct sequences of short programs to compose into long programs can still be challenging.
Therefore, we not only propose an architecture 
that can learn to hierarchically compose programs, denoted the Hierarchical Neural Program Synthesizer (HNPS),  
but also devise a training schema that allows for 
efficiently learning this hierarchical synthesizer.

To this end, we first construct a \textit{short program dataset}
that contains short programs with their corresponding input/output pairs.
Then, we learn a task embedding space 
that represents the space of short program behaviors, 
along with a program decoder that can 
decode a task embedding into a short program.
Next, we create a \textit{composed program dataset} 
in which the composed programs are generated by sequentially composing short programs. 
We then train a ``program composer'' on this dataset
% to comprehend the task specification (\eg input/output pairs) 
to produce a sequence of task embeddings, each representing a latent short program, which are decoded by the program decoder
into short programs and sequentially composed to yield the synthesized program. 
This training schema allows for 
generating programs via program-by-program composition rather than token-by-token composition, thus enabling additional supervision through 
the task embeddings 
instead of only maximizing the likelihood based on produced program tokens.
% We demonstrate that HNPS trained in this manner can synthesize unseen long programs which are not generated through short program composition.

% \sun{experiments}

To evaluate the proposed framework, 
we consider program synthesis in a string transformation domain, 
where a task specification consists of input/output strings 
(\eg $[\texttt{John-01}/\texttt{John}, \texttt{Mary-02}/\texttt{Mary}]]$) %\texttt{Bob-03}/\texttt{Bob}]]$)
and a program consists of a sequence of string manipulation operations
(\eg \texttt{SubStr(0, Regex("-", 0, Start))}),
similar to \citet{devlin2017robustfill, hong2020latent}.
Experiments show that
our framework is superior 
at synthesizing longer programs 
which induce more complex string manipulations.
Furthermore, we perform extensive ablation studies 
which justify the hierarchical architecture
and the latent supervision made possible 
by the proposed training schema and composed program dataset.

% A detailed discussion of related work can be found in~\mysecref{sec:related_work}.
% !TEX root = ../main.tex
\vspacesection{Related Work}
\label{sec:related_work}

%\Skip{
\myparagraph{Neural program induction} 
Program induction methods~\citep{graves2014neural, neelakantan2015neural, kaiser2015neural,reed2016neural,
gaunt2017differentiable, Penkov2017, xiao2018improving, lazaro2018beyond, pierrot2019learning, sun2020program} 
are designed to implicitly induce underlying programs 
to mimic the behaviors demonstrated in given task specifications.
%such as input/output pairs or expert demonstrations. 
To this end, most methods employ 
external memory~\citep{graves2014neural, zaremba2016learning},
modular frameworks with modularized supervision~\citep{reed2016neural, cai2017making, xu2018neural},
or sophisticated attention mechanisms~\citep{devlin2017neural}
to acquire programmatic behaviors.
In contrast, we are interested in explicitly synthesizing 
human-readable programs.
%}

\myparagraph{Neural program synthesis} 
Program synthesis
methods~\citep{balog2016deepcoder, bosnjak17a,parisotto2016neuro, devlin2017robustfill, bunel2018leveraging, shin2018improving, 
chen2019executionguided, liu2018learning,
sun2018neural, lin2018nl2bash,abolafia2018neural, verma2018programmatically, liao2019synthesizing, ellis2019write,ellis2020dreamcoder, gecco_program_synthesis_paper, hong2020latent, silver2020few, trivedi2021learning, liu2023hierarchical}
explicitly generate programs that can be executed to
perform the tasks from given specifications. 
These task specifications can range from strings of program input/output pairs to demonstrations of agent behaviors.
% or even a desired reward in a reinforcement learning setting~\citep{trivedi2021learning, verma2018programmatically}.
Compared to these works, we are particularly interested in
developing a framework that can synthesize longer programs with complex behaviors. Other methods, such as BUSTLE~\citep{odena_bustle_2021}, focus on incorporating neural methods into bottom-up search, combinatorially iterating through many potential programs. In contrast, HNPS synthesizes one program prediction.

Recently, pre-trained large language models (LLMs) have been employed for program synthesis from task specifications such as descriptions of desired program behavior or direct input/output examples~\citep{chen2021openaicodex, Nijkamp2022ACP, li2022alphacode}. These large models are trained on massive corpora of code which allow them to demonstrate impressive performance in synthesizing programs. However, they still generate tokens sequentially and can encounter the same issues as domain-specific token-by-token synthesis methods. In addition, transformer-based models~\citep{vaswani_attention_2017} have been adapted with novel program synthesis-tailored attention mechanisms~\citep{shi2022dl4cnps}. Our method is distinct in that it incorporates supervision against the latent embedding vectors in addition to the individual program tokens.
%- our ablation studies show the value of this supervision.
%with respect to the program aliasing problem and weak supervision on critical tokens. 
The training framework we propose is complementary to the advances made by LLMs and can be combined with 
such architectures and massive datasets in the future.

%!TEX root = ../main.tex

\vspacesection{Problem Formulation}
\label{sec:problem}

Our goal is to
synthesize programs from task specifications.
In this section, we formally describe our definition of programs and task specifications
as well as the specific problem formulation.

%\begin{figure}[t]
\begin{wrapfigure}[12]{R}{0.48\textwidth}
\centering
\vspace{-0.6cm}
\begin{mdframed}
\vspace{-0.2cm}
{%\footnotesize
 \dslfontsize
    \begin{align*}
    \text{Program}\ \rho &\coloneqq \text{Concat}\  (\ e_1\ e_2\ e_3 \ ...\ )\\
    \text{Expression} \ e &\coloneqq \text{ConstStr}\ (\ s \ )\ | \ \text{SubStr} \ ( \ q_1 \ q_2 \ ) \\ 
    \text{Position} \ q & \coloneqq \text{ConstPos}\ (\ k\ )\ | \ \text{Regex}\ (\ x\ n\ b\ )  \\
    \text{Boundary b}\ &\coloneqq \text{Start}\ | \ \text{End}\\
    \text{Regex} \ x &\coloneqq s\ | \ \text{Word}\ | \ \text{Num}\ | \ \text{Alphanum}\ | \ ... \ | \ \text{Char} \\
    \text{Position index} \ k & \coloneqq -20 \ | \ -19 | \ ... \ | \ 19 \ | \ 20\\
    \text{Regex index} \ n & \coloneqq -3 \ | \ -2 | \ ... \ | \ 2 \ | \ 3\\
    \text{Character}\ s &\coloneqq \text{"SPACE"}\ | \ \text{"."}\ | \ \text{","} \ | \ ...\ | \text{"?"}  \\ 
    \end{align*}
}
\vspace{-0.93cm}
\end{mdframed}
    \vspace{-0.25cm}
    \caption[]{
        \small
        The domain-specific language (DSL) for constructing string manipulation programs.  
        %A program consists of domain-dependent perceptions, actions, and control flows.
        \label{fig:dsl}
    }
\end{wrapfigure}

\noindent \textbf{Program and Domain Specific Language.} 
% \jesse{WE NEVER ACTUALLY JUSTIFY WHY WE USE STRING TRANSOFMRATION. NEED TO FIX HERE AND IN EXPERIMENTS SECTION. RELATE TO FLASHFILL, SPECIFICALLY MENTION FLASHFILL}
Programs considered in this work are constructed
based on
a Domain Specific Language (DSL) for string transformation tasks similar to DSLs for string transformation tasks used in (\citet{devlin2017robustfill,hong2020latent}).
The DSL establishes the possible space of valid programs 
as well as enables sampling programs.
An example of our DSL is shown in \myfig{fig:dsl}.
The DSL defines a set of operations 
(\eg \texttt{SubStr($k_1$, $k_2$)}, \texttt{Concat($e_1$, $e_2$, ...)}, 
\texttt{ConstStr($s$)})
and their parameters, such as numbers (\eg $k_1$, $k_2$) and strings (\eg $s$).
A program consists of a sequence of expressions which would output a sub-string of the input string or a constant string. The output of the expressions would then be concatenated to produce the final output string of the program.
For example, a program \texttt{Concat(SubStr(0, 1), ConstStr("!"))} 
takes an input string \texttt{John-01}
and produces an output string \texttt{J!}.
% \sun{do we want to talk about assuming random program and string generators?}

\myparagraph{Task Specification}
A task specification describes 
a desired user intent 
consisting of inputs and desired program execution results.
For example, $[\text{Input}: \texttt{October}, \text{Output}: \texttt{oct}]$
can serve as an input/output pair task specification for a string transformation program. Note that several different programs could satisfy this example specification, so in practice, multiple input/output pairs are usually specified to the program synthesis network. Importantly, the mappings between task specifications and programs are many-to-many---multiple task specifications can be used to describe the same desired program and many programs can satisfy the same task specification, signaling the difficulty of program synthesis.
In general, providing a task specification should be easier than asking non-expert users to write the program.
Commonly used representations for 
task specifications include
input/output (I/O) pairs\cite{devlin2017robustfill, hong2020latent},
demonstrations\cite{sun2018neural, trivedi2021learning,Penkov2017},
or natural language instructions\cite{li2022alphacode, Nijkamp2022ACP}.

\myparagraph{Problem Formulation} 
% \jesse{We consider developing a framework that can synthesize \textit{long} and \textit{complex} programs which satisfy given task specifications.}
We develop a framework that can 
synthesize a program from a task specification as described above.
In particular, we focus on long programs that 
can induce more complex execution results.
Specifically, we consider our string transformation domain,
where a task specification consists of a set of input/output string pairs
(\eg $[\texttt{John-01}/\texttt{John}, \texttt{Mary-02}/\texttt{Mary}, \texttt{Bob-03}/\texttt{Bob}]$)
and a program comprises a sequence of string manipulation operations
(\eg \texttt{SubStr(0, Regex("-", 0, Start))}). The long programs we consider would typically up to 63 tokens long and contain 2 to 4 ConstStr or SubStr expressions.
%!TEX root = ../main.tex

\vspacesection{Approach}
\label{sec:approach}

\begin{figure}[h!]
\centering
     \includegraphics[width=\textwidth]{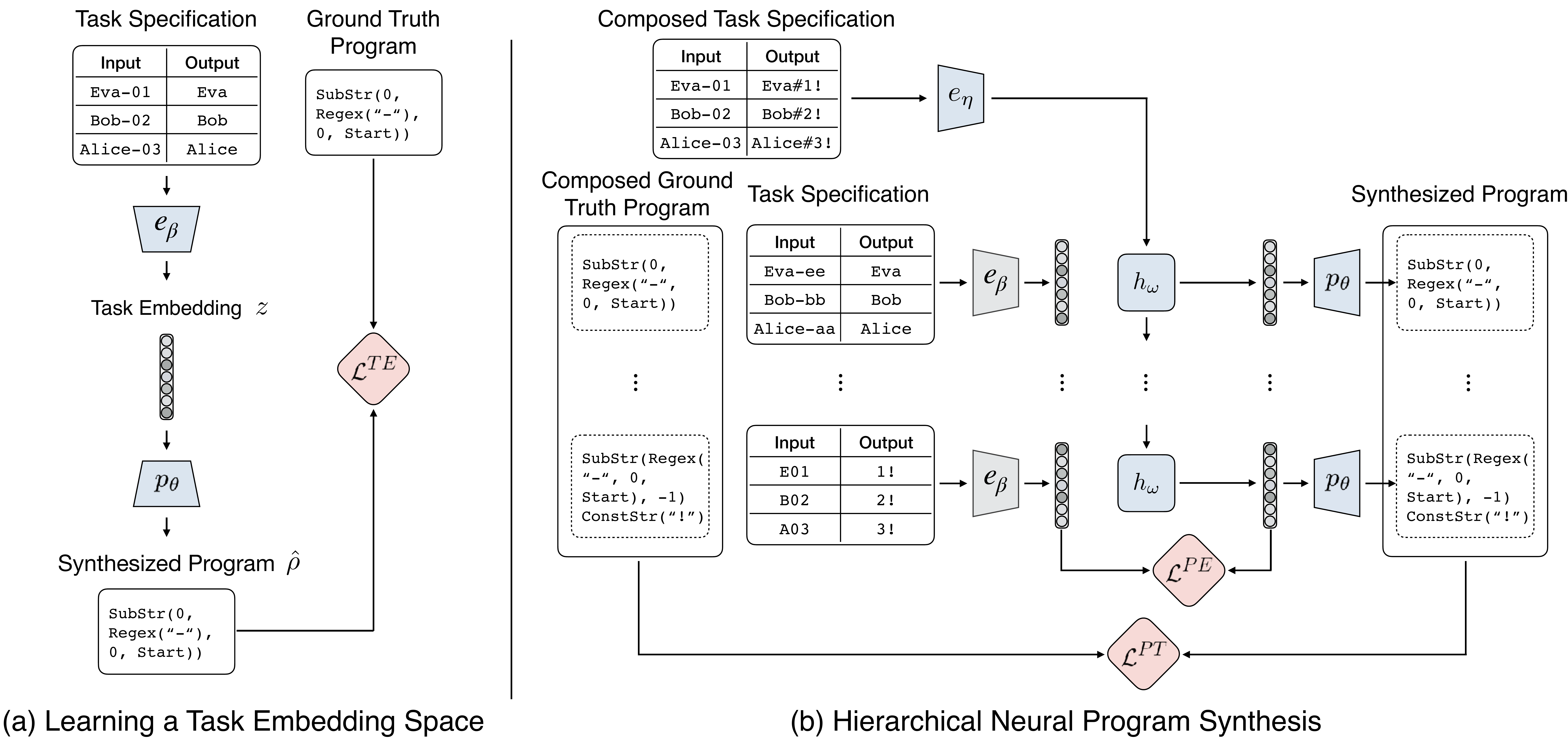}

    \caption[]{
        \small 
        \textbf{Framework Overview.}
        \textbf{(a)} Learning a Task Embedding Space: We first learn a task embedding space by training a task encoder, $e_\beta$, to encode a task specification (e.g. I/O pairs) $\sigma$ of a short program to a task embedding which is decoded by $p_\theta$ to a program $\hat{\rho}$. $e_\beta$ and $p_\theta$ are trained with task embedding loss $\mathcal{L}^{\text{TE}}_{\theta, \beta}$. 
        \textbf{(b)} Hierarchical Neural Program Synthesis: We then train a \emph{composed} task encoder $e_\eta$ initialized with the weights from the task encoder $e_\beta$ to encode the given I/O pairs from long programs to embeddings. The program composer $h_\omega$ uses this encoded input to continually predict task embeddings to be decoded by the pretrained $p_\theta$ decoder. These decoded programs are finally composed to form the complete synthesized program. The model is trained to synthesize long programs with a program token loss $\mathcal{L}^{\text{PT}}$ and match the ground truth program embeddings with the program embedding loss $\mathcal{L}^{\text{PE}}$.
        \label{fig:model}
    }
\end{figure}
% \jesse{a little repetitive now}
Our goal is to design a framework that can synthesize 
a long program given a task specification.
% Specifically, we are interested in learning to synthesize
% programs that are much longer than the ones studied in prior work.
To this end, 
we propose the Hierarchical Neural Program Synthesizer (HNPS), 
a framework that learns to compose programs 
to form longer programs, 
as illustrated in \myfig{fig:model}.
In the following, 
we describe how we learn a program embedding space 
in~\mysecref{sec:stage1} from short programs and simpler task specifications.
Then, we describe how we produce a dataset for learning 
to compose programs in~\mysecref{sec:approachdataset}.
Finally, 
we present 
% how we obtain supervision for learning to compose programs and 
how HNPS learns 
to synthesize a program given a task specification
by composing programs, 
described in~\mysecref{sec:hnps}.

\vspacesubsection{Learning a Task Embedding Space}
\label{sec:stage1}
To learn a task embedding space, 
we build a short program dataset $D_\text{short}$ by randomly sampling programs from the DSL, and we generate task specifications (\eg input/output pairs) by running sampled programs with randomly generated inputs.
We then train, on programs $\rho$ and corresponding task specifications $\sigma$ sampled from $D_\text{short}$, 
a neural program synthesis model that consists of a task encoder $e_\beta$ which encodes a task specification $\sigma$ to a task embedding, 
and a program decoder $p_\theta$ which synthesizes the program from the embedding. 
Both the task encoder $e_\beta$ and the program decoder $p_\theta$ 
are recurrent networks trained to 
optimize the token-by-token cross-entropy loss to maximize the likelihood of the 
ground truth program tokens. We denote this the task embedding loss, $\mathcal{L}_{\theta, \beta}^\text{TE}$:
\begin{equation}
\mathcal{L}_{\theta,\beta}^{\text{TE}}(\bm{\rho}) = -
\mathbb{E}_{(\bm{\sigma}, \bm{\rho}) \sim D_\text{short}} \left[\log
p_\theta\left(\boldsymbol{\rho}\vert e_\beta \left( \bm{\sigma} \right) \right) \right].
%+ \beta
%D_\text{KL}(q_\phi(\mathbf{z}\vert\boldsymbol{\rho})\|p_\theta(\mathbf{z})).
 \label{eq:ploss} 
 \end{equation}

Ideally, this learned embedding space can serve as a good representation space for desired program behavior, grouping similar programs and their task specifications closer together, and keeping dissimilar programs and their task specifications farther apart.

\vspacesubsection{Creating The Composed Long Program Dataset}
\label{sec:approachdataset}

Our ultimate goal is to learn to compose short programs to form a longer program.
Therefore, 
% we take advantage of the inductive bias that each long program is composed of independent short programs,
we create a dataset $D_\text{composed}$ which consists of 
long programs obtained by composing shorter programs 
from the program dataset $D_\text{short}$. 
We denote these composed long programs $\bar{\rho} = \left[ \rho_1,...,\rho_N \right]$, 
where each $\rho_i$ is a short program. 
Task specifications for each composed program is randomly sampled using the technique detailed in \ref{sec:dataset}
Such a composed program dataset will give us access 
to ground truth information of how each long program 
is composed and allow the model to learn program composition.
For all long programs in $D_\text{composed}$, we also include the task embeddings of their sub-programs, calculated with the task encoder $e_\beta$ using the task specifications of short programs from $D_\text{short}$. The information from task embeddings will enable us to train our model to fully utilize our learned task embedding space.

\vspacesubsection{Hierarchical Neural Program Synthesis}
\label{sec:hnps}

% Intuitively, synthesizing long programs from scratch by sequentially predicting  
% program tokens becomes more difficult as the sequence length increases, a hypothesis 
% that we empirically validate in our experiments (see~\mysecref{sec:results}).
% Therefore, 
To synthesize a long program, program-by-program 
instead of token-by-token,
we propose to leverage the program decoder $p_\theta$
learned in~\mysecref{sec:stage1}
as a low-level module to produce short programs.
Then, we employ a \textit{program composer} $h_\omega$
that produces a sequence of task embeddings which 
can be decoded into short programs 
and sequentially composed to form a long program.\footnote{We acknowledge the fact that our model is limited to linear compostion of sub-programs. However, we believe specific modes of composition is orthogonal to our main focus of introducing heirchical composition as a solution to scalable program synthesis. We hope to extend to more general model of compostion in future work.}

Specifically, our Hierarchical Neural Program Synthesizer consists of the following three components. 
A \emph{composed} task encoder $e_\eta$ learns to encode given task specifications from long programs to embeddings
(\ie string I/O pairs). Since task specifications from $D_\text{short}$ and $D_\text{composed}$ are similar, parameters from the task encoder $e_\beta$ can be used as initialization for the composed task encoder $e_\eta$.
Then, the program composer $h_\omega$ 
takes the output of the composed task encoder as input and
sequentially produces task embeddings until
a long program that satisfies the task specification is generated (in the case of string transformaiton, when the program can generate correct output from all given input strings)
, or until the maximum program length is reached. Finally, the program decoder $p_\theta$ trained during the construction of the task embedding space will generate the corresponding short program. Since predicting the exact task embedding can be difficult, the program decoder $p_\theta$ is fine-tuned along with other components and the task embedding space is allowed to drift.
To train the Hierarchical Neural Program Synthesizer,
we propose to optimize the following objectives.

\myparagraph{Program Token Loss} 
The program token loss $\mathcal{L}^{\text{PT}}$ 
simply aims to maximize the log likelihood of the ground truth program tokens
in the synthesized program via the cross-entropy loss. 
After sampling a composed program $\bar{\rho}$ and its task specifications $\sigma$, we apply the cross-entropy loss to maximize the log likelihood of each individual short program $\rho_1...\rho_N$ that makes up $\bar{\rho}$.
%the task specifications are first encoded by the behavior encoder $e_\eta$, then composed by the program composer $h_\omega$, and decoded into program tokens by the program decoder $p_\theta$. We apply the 
This loss is propagated throughout all three networks. Note that $h_\omega$ is
a recurrent network so its predictions depend on earlier short programs; therefore, losses from later shorter programs will be backpropagated through time to update both $e_\eta$ and $h_\omega$ to improve overall reconstruction of the entire long program:
\begin{equation}
\mathcal{L}^{\text{PT}}_{\theta, \omega, \eta} = -\mathbb{E}_{(\bm{\bm{\sigma}, \bar{\rho})} \sim D_\text{composed}} \left[ \frac{1}{N}\sum_{i=1}^N \log p_\theta \left(\bm{\rho_i} \vert h_\omega \left(e_\eta\left(\bm{\sigma}\right)\right)\right)  \right].
\end{equation}

\myparagraph{Program Embedding Loss} 
%Given that the program composer is initialized from scratch,
%it is difficult for it to know how to output sensible behavior embeddings
%that can be decoded by $p_\theta$. 
Even though the task embedding space may change as $p_\theta$ is fine-tuned, embeddings from the initial task embedding space still provide useful grounding for the program composer to output sensible task embeddings.
To provide regularization on the output space of the program composer,
%we decompose each composed training program into a set of shorter programs.
we embed the task specifications $\rho_i$ of each short program that composes longer programs in $D_\text{composed}$ with the original fixed task encoder $e_\beta$.
%we obtain an embedding of each short program that composes each using 
%the learned program encoder $q_\phi$.
Then, 
we apply a program embedding loss $\mathcal{L}^{\text{PE}}$ that
aims to minimize the Euclidean distance between 
this ground truth (\ie encoded) program embedding
and the program embedding produced by the program composer, for all $N$ subprograms in the composed program:
\begin{equation}
\mathcal{L}^{\text{PE}}_{\eta, \omega} = \mathbb{E}_{(\bm{\sigma}, \bm{\sigma_1},...,\bm{\sigma_N})\sim D_{\text{composed}}}\left[ \frac{1}{N} \sum_{i=1}^N|| h_\omega\left(e_\eta\left(\bm{\sigma}\right)\right) - e_\beta \left( \bm{\sigma_i} \right)||_2 \right] .
\end{equation}

In our experiments in \mysecref{sec:results}, we find that this loss
is especially useful in scenarios with less training data to ground the program composer to the space of sensible program embeddings.

%\myparagraph{Termination Loss} 
%To determine when to cease to predicting program embeddings,
%we employ a termination loss $\mathcal{L}^{\text{Term}}$ to
%train the program composer
%using a binary Cross-Entropy loss.

%\begin{equation}
    %\text{Termination loss here}
%\end{equation}

In summary, we propose the following objective
for learning to synthesize a program by 
processing given task specifications and composing programs:
$
    \min_{\eta, \omega, \theta} 
    \mathcal{L}^{\text{PT}} + 
    \lambda \mathcal{L}^{\text{PE}}
$
% where we perform alternating supervised and RL loss updates.
where $\lambda$
controls the importance of each loss.
Note that the program embedding loss
comes from our contribution of 
composing a dataset with shorter programs 
to construct a dataset with decomposable longer programs.
%!TEX root = ../main.tex

\vspacesection{Experiments}
\label{sec:experiment}

We evaluate our framework on datasets generated on the aforementioned string transformation task, detailed in Section~\ref{sec:dataset}. We then justify the design decisions of \method\\ by comparing against baselines and ablations detailed in Section~\ref{sec:baselines}. Finally, we present the results of our experiments in Section~\ref{sec:results}.

\vspacesubsection{Datasets}
\label{sec:dataset} % temp location just so we can see the table
We create two datasets, $D_\text{short}$ and $D_\text{composed}$, where $D_\text{composed}$ consists of long programs which are obtained by composing shorter programs from $D_\text{short}$. String I/O pairs are created along with their corresponding programs during the generation process. $D_\text{short}$ has 100,000 programs, each of which contains only one ConstStr or SubStr expression. We create these short programs by randomly sampling from our DSL in a manner similar to that of prior work \citep{trivedi2021learning}. For each sampled program, we randomly generate 1,000 strings as potential input strings, and the first 20 input strings that can be executed without exceptions are stored in our dataset along with their corresponding output strings. If less than 20 input strings in the 1,000 generated are executed without exceptions for a program, then that program is discarded. We repeat this process until the target number of programs are generated and each with 20 I/O pairs as task specification.  For $D_\text{composed}$, we take advantage of the modality of our DSL to by linearly composing expressions from short programs into long programs. We generate 200,000 composed programs by composing 2-4 randomly sampled programs from $D_\text{short}$. Program composition process is detailed in the appendix. For each composed program in $D_\text{composed}$, the same method is used to generate I/O pairs. 
We split each dataset into training and validation sets to be able to evaluate model generalization performance during training.

Finally, we generate a long program dataset, $D_\text{long}$, which consists of 30,000 programs, each containing 2-4 expressions (resulting in programs up to 63 tokens long) by directly sampling from our DSL. Notably, $D_\text{long}$ is not constructed by composing short programs.
% in which we do not assume that the decomposition of its long programs into shorter ones are given. 
% ( may be a problem to say this since each short program is only one expression long, so decomposition is still possible by simply separate each expressions.
We evaluate the performance of \method\\ on this testing dataset to evaluate its ability to synthesize unseen, long programs.

\Skip{
\vspacesubsection{Domain}

To evaluate the proposed framework, 
we consider the Karel domain~\citep{pattis1981karel}, 
as featured in~\cite{bunel2018leveraging, shin2018improving, sun2018neural},
which features an agent navigating through a gridworld with walls 
and interacting with markers. 
The agent has 5 actions for moving 
and interacting with marker 
and 5 perceptions for detecting obstacles and markers. 
The tasks of interest are shown in~\myfig{fig:karel envs}. Note that most tasks have randomly sampled agent, wall, marker, and/or goal configurations. When either training or evaluating, we randomly sample initial configurations upon every episode reset.
More details can be found in~\mysecref{sec:env details}.
}

\vspacesubsection{Baseline and Ablations}
\label{sec:baselines}
We evaluate \method\\ against baselines and its variants. 
The following baselines represent the family of neural synthesis methods that are designed to 
synthesize a program in a token-by-token manner.
\begin{itemize}
    \item \textbf{Na\"{i}ve}: A na\"{i}ve synthesis baseline trained on $D_\text{composed}$ which encodes the I/O pair tokens to produce an embedding which is used to synthesize the full-length program. 
    % Both the I/O encoding network and the synthesis network are LSTM networks.
    This baseline is expected to struggle at learning to synthesize long programs with complex behaviors from scratch.
    \item \textbf{Na\"{i}ve-short}: Na\"{i}ve but the model is only trained on $D_\text{short}$. 
    This baseline is expected to learn well from $D_\text{short}$ but should zero-shot generalize poorly
    to $D_\text{long}$.
    \item \textbf{Na\"{i}ve-short-finetune}: Na\"{i}ve-short but the model is finetuned on $D_\text{composed}$. 
    This baseline utilizes both $D_\text{short}$ and $D_\text{composed}$ in the same order as our proposed method. The difference is that it has a flat architecture without hierarchy like ours.
    The performance gap between this baseline and ours 
    should justify the hierarchical design of our proposed framework.
    %\item LLM: We finetune a large language model (LLM), partially trained on publicly available Github code, on our training dataset.
\end{itemize}    

We also include a state-of-the-art search-based program synthesis baseline~\citep{odena_bustle_2021}.
\begin{itemize}
   \item \textbf{BUSTLE}: A bottom-up search technique, designed for tasks like string manipulation, that combinatorially iterates through the program DSL to synthesize a program that meets the task specification. To execute the search in a more informed way, the method ranks certain sub-expressions to be searched before other sub-expressions. 
   %based on string similarity heuristics. 
   Runtime limit is set to 500,000 expressions.
   %\item \textbf{BUSTLE-NN}: BUSTLE-heuristic guided by a neural network trained to predict whether an intermediate program sub-expression is likely to be used in the final result.
\end{itemize} 

To justify our design choices, the following variants of \method\\ are considered for ablation studies.

\begin{itemize}    
    \item \textbf{H-Na\"{i}ve-PT}: A na\"{i}ve \text{hierarchical} synthesis baseline in which the decoder is trained from scratch (\ie we skip the learning embedding space stage), and only the program token loss $\mathcal{L}^{\text{PT}}$ is applied. 
    This ablation learns to compose short programs to create longer ones, 
    except that it does not learn in a two-stage fashion.
    The performance gap between this ablation and our proposed framework
    should justify the importance of learning a task embedding space.
    \item \textbf{\method\\-PT}: An ablation in which only the program token loss $\mathcal{L}^{\text{PT}}$, 
    which still provides strong supervision for relatively shorter programs, is applied.
    This ablation first learns a task embedding space from 
    $D_\text{short}$ and then learns from $D_\text{composed}$ while optimizing $\mathcal{L}^{\text{PT}}$. Here, we seek to analyze in which contexts the absence of the program embedding loss may be more or less detrimental to execution accuracy performance, given that ground truth task embeddings can be noisy.
    \item \textbf{\method\\ (\method\\-PT+PE)}: The full \method\\ method. 
    It first learns a task embedding space from 
    $D_\text{short}$. 
    Then, it optimizes both losses (the program token loss $\mathcal{L}^{\text{PT}}$ and the program embedding loss $\mathcal{L}^{\text{PE}}$) while learning from $D_\text{composed}$. 
\end{itemize}

\vspacesubsection{Results}
\label{sec:results}
We present results of HNPS and comparison methods on synthesizing programs 
in both an unseen, non-composed, long program dataset and in unseen programs in test set of $D_\text{composed}$ in \mysecref{sec:main_results}.
Then, we analyze the task embedding space learned by HNPS and how it aids
long program synthesis in \mysecref{sec:embedding_exp}. Finally, we carefully examine
the effect of the program embedding loss on synthesizing programs of different lengths and with different dataset sizes in \mysecref{sec:ablation_pe}.

\subsubsection{Execution Accuracy on Long Programs}
\label{sec:main_results}

%\begin{wraptable}{r}{0.55\textwidth}
\begin{table}[h!]%[t]
    \centering
    \small
    \caption{\small 
    \textbf{Performance on Unseen Test Programs.}
    We evaluate execution accuracy over \textit{unseen} test programs in our dataset $D_\text{long}$, where execution accuracy refers to the percent of synthesized programs that produce the correct output for all given input strings. The x-axis represents the length of the ground truth program (number of program tokens).
    }
    \vspace{-0.3cm}
    \scalebox{0.85}{
    \begin{tabular*}{\linewidth}{l@{\extracolsep{\fill}}cccc}
        \toprule
        & \multicolumn{4}{c}{\textbf{Number of program tokens}} \\
        \cmidrule(lr){2-5}
        \textbf{Method} & \textbf{10-25} & \textbf{25-40} & \textbf{40-55} & \textbf{55-70} \\
        % & Program & Execution & Program & Execution \\

        \midrule
        Na\"{i}ve & 42.99\% &  11.88\%&  0.61\%& 0.01\% \\
        Na\"{i}ve-short & 0.00\% &  0.00\%&  0.00\%& 0.00\% \\
        Na\"{i}ve-short-finetune & 42.68\% &  16.75\%&  1.93\%& 0.02\% \\
        BUSTLE & 10.72\% & 0.13\% & 0.00\% & 0.00\% \\
    
        \midrule
        H-Na\"{i}ve-PT & 60.78\%&  27.12\%&  5.99\%& 0.41\% \\

        \midrule
        HNPS-PT & \textbf{65.76\%} & \textbf{32.49\%} & 9.4\% & 0.78\% \\
        HNPS (ours-full) & 64.25\%&  32.22\%& \textbf{9.71\%}& \textbf{0.87\%} \\
        \bottomrule
    \end{tabular*}
	}
    \label{table:main}
%\end{wraptable}    
\end{table}
\myparagraph{Evaluation on Non-Composed Long Programs} 
We evaluate execution accuracy over \textit{unseen} test programs in our dataset $D_\text{long}$, where execution accuracy refers to the percent of synthesized programs that produce the correct output for all given input strings.
\begin{wrapfigure}[17]{R}{0.45\textwidth}
\vspace{-0.5cm}
\centering
     \includegraphics[width=0.45\textwidth]{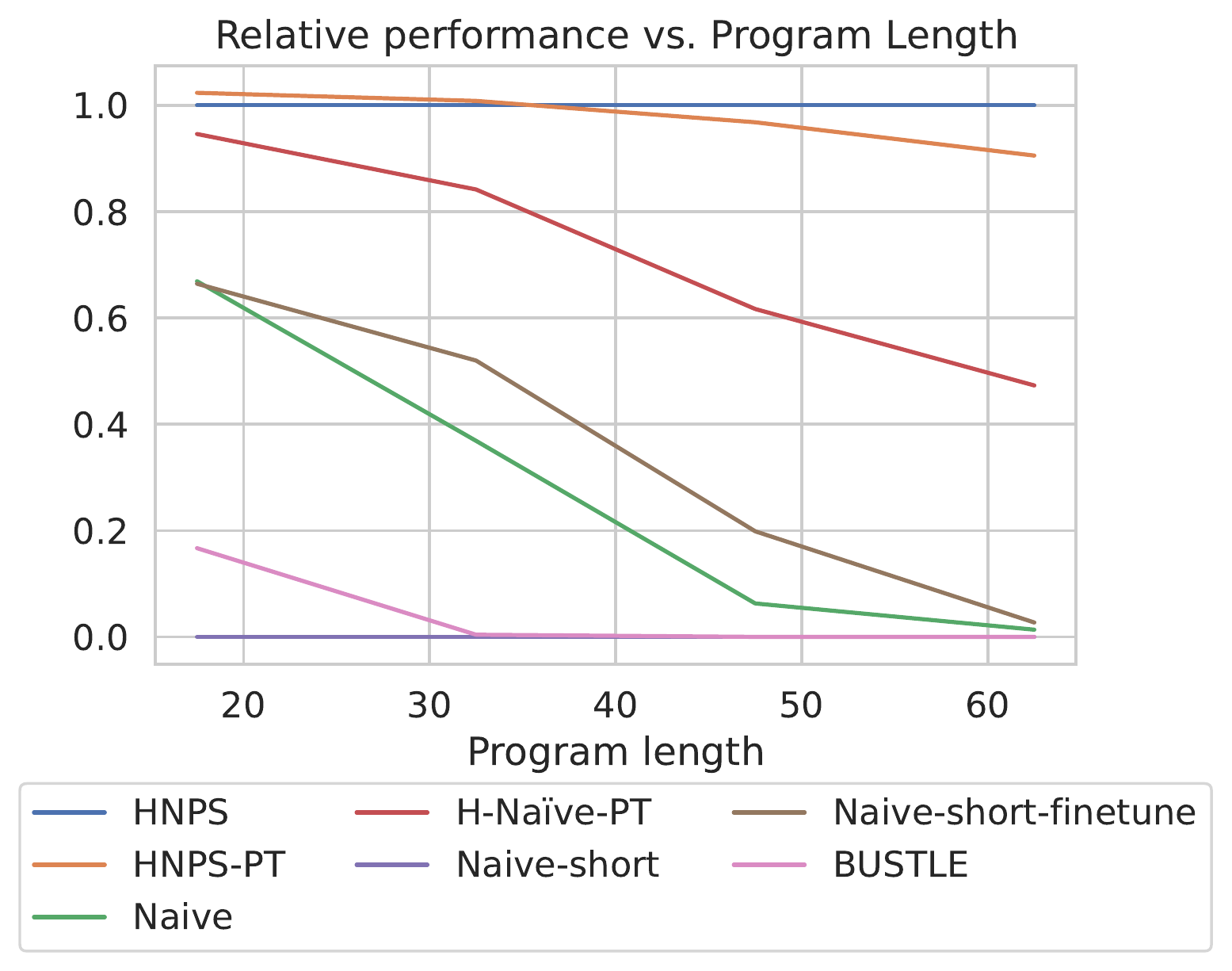}
    \vspace{-0.5cm}
    \caption[]{
        \small 
        Relative performance of all baselines and HNPS evaluated on $D_\text{long}$. We normalize the execution accuracy of each method by scaling the accuracy of HNPS to 1.0.
        \label{fig:relative_performance}
    }
\end{wrapfigure}

Results from our experiments are shown in Table~\ref{table:main}. We additionally provide a visualization of relative performance on different program lengths in Figure~\ref{fig:relative_performance}, where we normalize the execution accuracy of each method by scaling the accuracy of HNPS to 1.0. For programs longer than 40 tokens, our method is able to achieve better performance than the baselines and ablations with 9.71\% execution accuracy on programs 40-55 tokens long and 0.87\% on programs 55-70 tokens long. For programs shorter than 40 tokens, HNPS-PT, where program embedding loss is not applied, yields the best performance with 65.76\% execution accuracy on programs with 10-25 tokens and 32.49\% on program with 25-40 tokens. We further analyze performance differences between HNPS-PT and HNPS in shorter programs in later sections. 

With only the hierarchical architecture and without the pretrained task embedding space, H-Na\"{i}ve-PT is still able to outperform other non-hierarchical methods. However, its relative performance compared to HNPS drops significantly as program length increases, reaching less than 50\% of HNPS's execution accuracy on 55-70 token programs. This demonstrates that the task embedding space is crucial in training hierarchical models.

The non-hierarchical baselines Na\"{i}ve, Na\"{i}ve-short, and Na\"{i}ve-short-finetune perform significantly worse than our method. All three baselines have close to zero execution accuracy on 55-70 token programs, even for Na\"{i}ve-short-finetune which finetunes the same task encoder and program decoder as HNPS on the same datasets. This lends credence to our claim that the token-by-token generation scheme is fundamentally limiting to a program synthesis model's ability to scale to longer programs.

The search based method, BUSTLE, also performs poorly, achieving only 10.72\% and 0.13\% execution accuracy on programs with 10-25 and 25-40 tokens, respectively. For programs greater than 40 tokens, BUSTLE is never able to synthesize a program that completely meets the task specification. Longer program synthesis settings, like the one we evaluate on, seem to be too complex for bottom-up search methods. Even though it is able to prioritize its search towards sub-expressions that have a higher likelihood to appear in the final solution, BUSTLE is unable to efficiently search through the large space of possible programs and yield a solution.

\myparagraph{Evaluation on Composed Long Programs} 
We evaluate execution accuracy of all methods over training programs and \textit{unseen} test programs in our $D_\text{composed}$ dataset. Exact results are detailed in appendix Section~\ref{sec:d_composed_eval}. Similar to the results from evaluation on $D_\text{long}$, HNPS \& HNPS-PT are able to outperform H-Na\"{i}ve-PT, which further highlights the importance of constructing a task embedding space. We also observe that non-hierarchical baselines (Na\"{i}ve-*) %(Na\"{i}ve, Na\"{i}ve-short, and Na\"{i}ve-short-finetune) 
have significantly lower performance than hierarchical methods on both training and test programs. This supports our assertion that without the hierarchical architecture, the model's ability to fit to a distribution of long programs is fundamentally limited. 

We also analyze the ``generalization gap,'' \ie the difference between training and test set performance, on $D_\text{composed}$. HNPS has a smaller generalization gap than both HNPS-PT and H-Na\"{i}ve-PT. This implies that the program embedding loss can help prevent overfitting in hierarchical models. This may be due to the fact that the program decoder learns to synthesize a diverse set of short programs during $D_\text{short}$ training. When training on $D_\text{composed}$, the embedding loss grounds the decoder to the original task embedding space, encouraging it to not overfit to only the short programs needed for the $D_\text{composed}$ training set. Therefore, the program embedding loss not only helps the composer to create longer programs (which we will explore further in Section~\ref{sec:embedding_exp}) but also helps ensuring the model does not overfit to the subset of short programs that are used to compose $D_\text{composed}$.

\begin{figure}[t]
\centering
    \caption[]{
        \small
        \textbf{Qualitative Examples.} We show an example of a program generated by HNPS, HNPS-PT, and Na\"{i}ve-short-finetune compared with the ground truth program. Non-critical errors, where the changes in tokens do not effect program behavior, are marked \textcolor{orange}{orange}. Critical errors, where the changes in tokens result in different program behavior, are marked \textcolor{red}{red}.
        %A program consists of domain-dependent perceptions, actions, and control flows.
        \label{fig:gen_prog}
    }
    \vspace{-0.25cm}
    \includegraphics[width=0.98\textwidth]{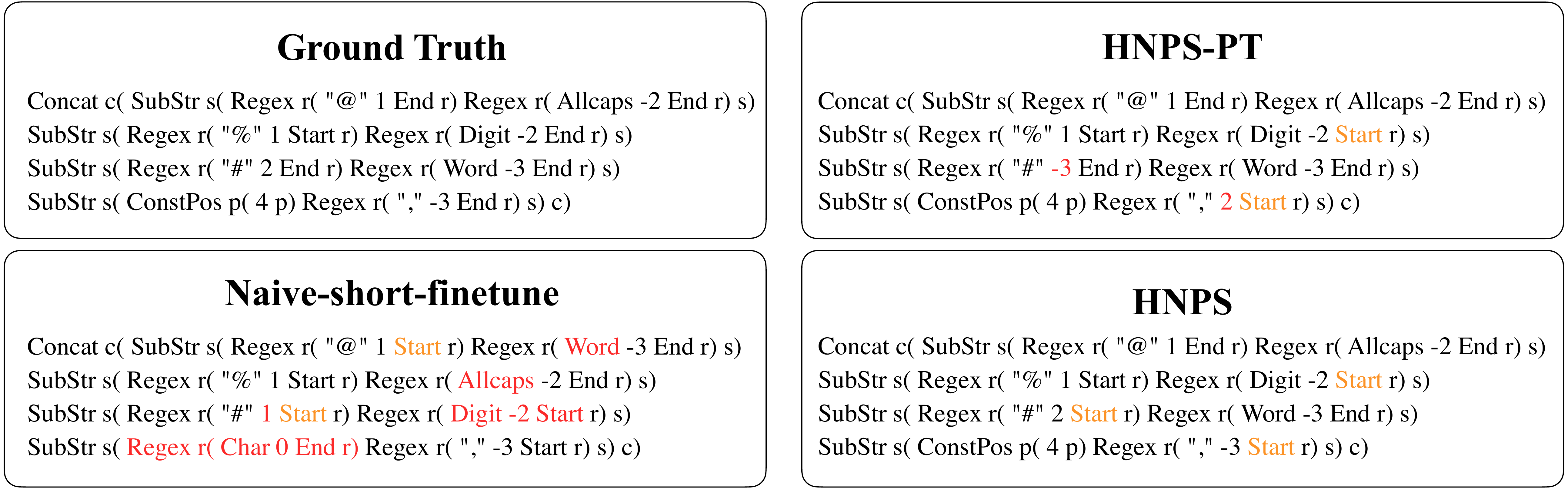}
\end{figure}

\myparagraph{Qualitative Examples} In Figure ~\ref{fig:gen_prog}, we show an example of programs generated by HNPS, HNPS-PT, and Na\"{i}ve-short-finetune compared with the ground truth program. To assess our method's ability to create a meaningful embedding space for program composition, we first introduce two concepts: critical/non-critical tokens and critical/non-critical errors. We define critical tokens as program tokens that if changed, will alter the behavior of the program and thus produce a different program output for a given string input. Following this, critical errors are characterized by errors of the program synthesis module on critical tokens. Conversely, non-critical tokens are those that can be replaced without affecting program behavior, and non-critical errors are mistakes on non-critical tokens. In the figure, we mark all non-critical errors \textcolor{orange}{orange} and all critical errors \textcolor{red}{red}. 

In the program generated by HNPS, there are 3 tokens that are different from the ground truth program, all of them confusing the "End" token with the "Start" token. However, even with these incorrect tokens, the HNPS-generated program still achieves correct behavior. This is because the DSL syntax dictates that "Regex" expressions that only capture one string token will refer to the same index position regardless of whether the "Start" or "End" token is specified. So, even with three token errors, the output of HNPS is able to satisfy the I/O specification. 

HNPS-PT's generated program has 4 token errors, which is just one more error than HNPS. Yet, 2 of the 4 errors made by HNPS-PT are in critical parameters of the "Regex" function that completely change the function output. This example correspond to a trend we observed in generated programs where HNPS tends to make same amount or more token errors compared to HNPS-PT, but HNPS makes less error in critical tokens, resulting in better execution accuracy. We suspect that this is because supervision for HNPS-PT on critical tokens is weak without the program embedding loss, resulting in more critical errors. 

For Na\"{i}ve-short-finetune, the generated program has several errors in regex pattern parameters and even a wrong function. This demonstrates that even though Na\"{i}ve-short-finetune is trained on the same datasets, its performance is still limited by the token-by-token prediction scheme, making it unable to recognize correct program behavior and synthesize accurate programs. 

% MOVE TO APPENDIX
\subsubsection{Analyzing the Task Embedding Space}
\begin{wrapfigure}[17]{R}{0.45\textwidth}
\centering
\vspace{-0.5cm}
     \includegraphics[width=0.45\textwidth]{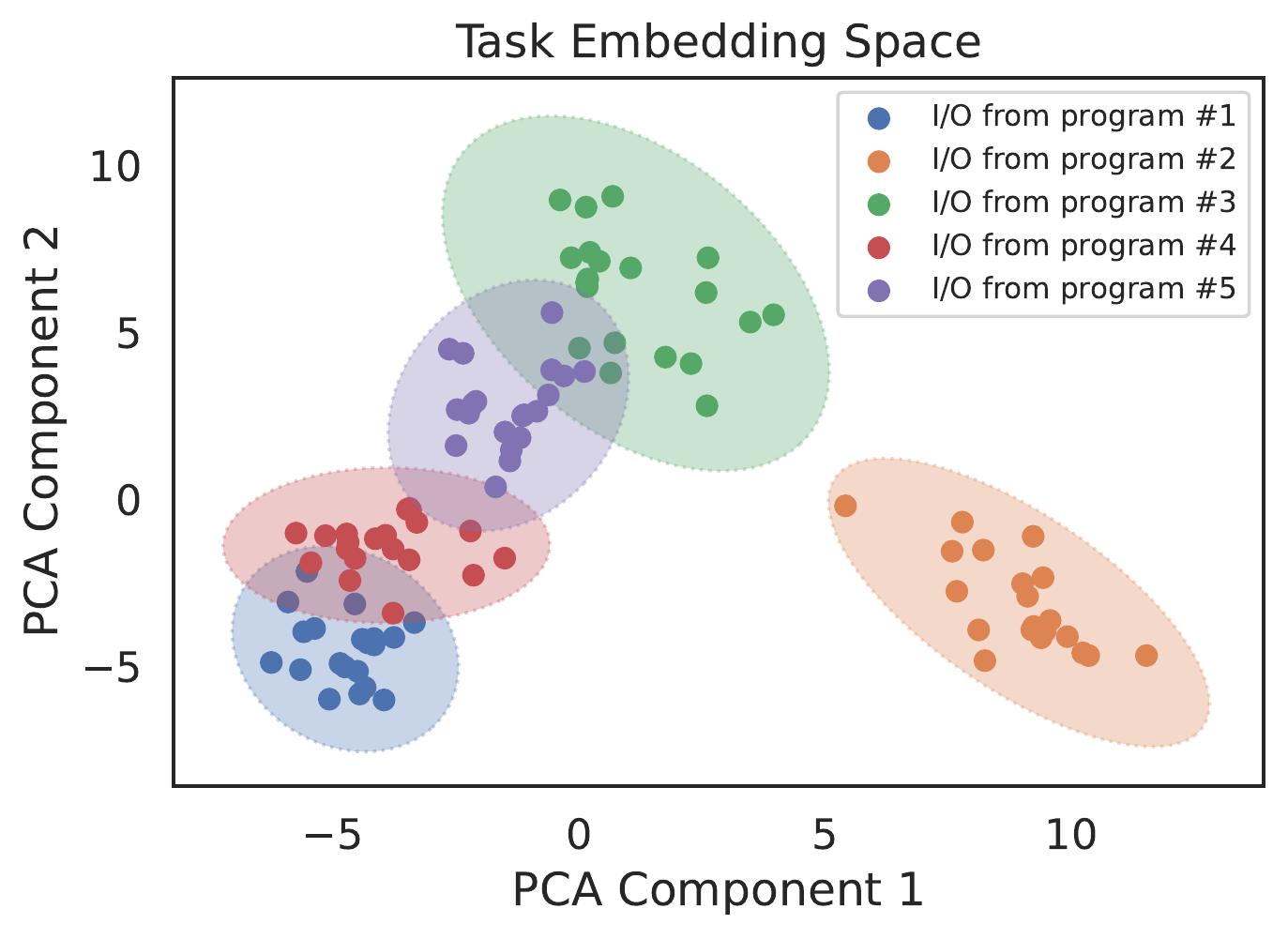}
    \vspace{-0.5cm}
    \caption[]{
        \small 
        We perform dimensionality reduction with PCA on the task embeddings of 5 programs randomly drawn from the test set of $D_\text{short}$. Each dot represents the embedding of an I/O pair and each color represent a different program. 
        \label{fig:embedding_space}
    }
\end{wrapfigure}
\label{sec:embedding_exp}
%\jesse{summarize the insights up front: say what we want the readers to believe, then prove it to them} 
We now analyze how learning the task embedding space aids 
with hierarchical program synthesis.
In Figure~\ref{fig:embedding_space}, we provide a visualization of the tasks embeddings of 5 programs randomly drawn from the test set of $D_\text{short}$. Dimensionality reduction is performed with principal component analysis (PCA) in order to project the task embeddings to a 2D space. We represent one embedding of an I/O pair with a dot in the figure and represent different programs using various colors. As shown in the figure, I/O pairs from the same program are clustered together and I/O pairs from different programs are separable, demonstrating that a meaningful embedding space is able to be learned and leveraged in our training procedure. 

We are also interested in whether our embedding space can identify ``critical tokens,'' tokens that would greatly change program behavior if swapped out for a different one. If nearly identical programs with minor critical changes are embedded farther apart, then our program composer is more resilient to the critical token problem that can affect standard neural synthesis methods.
To do so, we sampled 100 programs from $D_\text{short}$. For each sampled program $\rho$, we either make a change to one non-critical token to create $\rho_\text{non-critical}$, or change one critical token to create $\rho_\text{critical}$. To get the task embedding of $\rho_\text{non-critical}$ and $\rho_\text{critical}$, we generate 20 I/O pairs for each program and pass them through our task encoder $e_\beta$. In Table~\ref{table:emb_distance}, we show that the distance between embeddings of programs with a difference in a critical token is significantly larger than the distance between programs with a difference in one non-critical token. We thus confirm that our embedding space captures program semantics, as changes in critical tokens alter the behavior of the program more than changes in non-critical tokens, and our embedding space accordingly places the embedding of $\rho_\text{critical}$ further from $\rho$ than it does for the embedding of $\rho_\text{non-critical}$. We hypothesize that this smooth embedding space enables the program composer to more easily synthesize programs that match given task specifications.
\begin{wraptable}{R}{0.50\textwidth}
%\begin{table*}[t]

    \centering
    \small
    \caption{\small 
    \textbf{Embedding Distance with Changes in Tokens.}
    We sampled 100 programs each from the training set and test set of $D_\text{short}$ to compare the average euclidean distance between task embeddings of programs with a difference in one critical token and programs with a difference in one non-critical token.
    }
    \scalebox{0.9}{
    \begin{tabular*}{\linewidth}{lcc}
    \toprule
    difference type & train & test \\
    \midrule
    non-critical token & 2.92 $\pm$ 0.58 & 2.99 $\pm$ 0.63 \\
    critical token & 6.71 $\pm$ 1.98 & 10.50 $\pm$ 1.79 \\

    \bottomrule
    \end{tabular*}
	}
    \label{table:emb_distance}
\end{wraptable}    
% \end{table*}
\subsubsection{Ablating the Program Embedding Loss}
\label{sec:ablation_pe}
Now we analyze the effect of the program embedding loss specifically. 
In Table \ref{table:program_bembeeding_loss} we compare HNPS, our full method, and HNPS-PT, our method without the program embedding loss $\mathcal{L}^{\text{PE}}$, on $D_\text{long}$ as a function of the size of the composed program dataset. We examine these two methods on: (1) our full $D_\text{composed}$ containing 200,000 programs, (2) 50\% of $D_\text{composed}$, and (3) 25\% of $D_\text{composed}$. When trained with the full $D_\text{composed}$, HNPS only performs better than HNPS-PT on programs longer than 40 tokens. However, when using 50\% or 25\% of the data, HNPS's performance become significantly better than HNPS-PT across all program lengths. Especially for for programs longer than 55 tokens on 100k dataset, HNPS's execution accuracy is 0.16\% which is 8 times the performance of HNPS-PT. This suggests that the supervision from program embedding loss is particularly effective when data is scarce. We hypothesize that this is because in practice, ground truth task embeddings tend to be noisy due to the imperfect task encoder. When a good support for the program space is provided by training data, the model can learn to compose short programs from the program token loss alone, making a noisy program embedding loss less useful as supervision. However, when training data is scarce, learning from the token loss alone is difficult and it will need the information provided by program embedding loss to output sensible task embeddings despite the noisiness of the embedding loss.
%\begin{wraptable}[16]{r}{0.65\textwidth}
\begin{table*}[h!]%[t]
    \centering
    \small
    \caption{\small 
    \textbf{Effect of Dataset Size on Program Embedding Loss.}
    We examine HNPS and HNPS-PT on: (1) our full $D_\text{composed}$ containing 200,000 programs, (2) 50\% of $D_\text{composed}$ (100k), and (3) 25\% of $D_\text{composed}$ (50k). We report their execution accuracy on $D_\text{long}$.
    }
    \resizebox{0.9\textwidth}{!}{
    \begin{tabular*}{\linewidth}{ll@{\extracolsep{\fill}}cccc}
    \toprule
    & & \multicolumn{4}{c}{\textbf{Number of program tokens}}\\
    \cmidrule(lr){3-6}
    \textbf{Dataset Size} & 
    \textbf{Method} & \textbf{10-25} & \textbf{25-40} & \textbf{40-55} & \textbf{55-70} \\
    
    \midrule
    \multirow{2}{*}{200k}
    & HNPS & 64.25\% & 32.22\% & \textbf{9.71\%} & \textbf{0.87\%} \\
    & HNPS-PT & \textbf{65.76\%} & \textbf{32.49\%} & 9.4\% & 0.78\% \\
    
    \midrule
    \multirow{2}{*}{100k}
    & HNPS & \textbf{55.20\%} & \textbf{20.06\%} & \textbf{2.32\%} & \textbf{0.05\%} \\
    & HNPS-PT & 50.5\% & 14.55\% & 0.67\% & 0.01\% \\
    
    \midrule
    \multirow{2}{*}{50k}
    & HNPS & \textbf{42.23\%} & \textbf{7.71\%} & \textbf{0.12\%} & 0.00\% \\
    & HNPS-PT & 40.72\% & 4.5\% & 0.01\% & 0.00\%  \\
    
    \bottomrule
    \end{tabular*}
	}
    \label{table:program_bembeeding_loss}
%\end{wraptable}    
\end{table*}

% \myparagraph{Ablation study: non-composed programs} We present the results of our experiments in Table~\ref{table:main}. 
% We evaluate both program and execution accuracy over unseen test programs in our two datasets $D_\text{short}$ and $D_\text{long}$, where program accuracy refers to
% the percentage of synthesized programs which exactly match the ground truth program given the I/O pairs, and execution accuracy refers to the percent of outputs of synthesized programs that matches the expected outputs on given input-output pairs.

%!TEX root = ../main.tex

\vspacesection{Discussion}
\label{sec:conclusion}

In this work,
we study learning to synthesize programs from task specifications such as input/output pairs (I/O pairs).
In particular, we are interested in scaling up current 
neural program synthesis methods to synthesize long programs with more complex behaviors.
To this end, we propose the Hierarchical Neural Program Synthesis (HNPS) framework;
instead of producing a program token-by-token like most existing methods,
we propose to synthesize a program subprogram-by-subprogram.
Specifically, we first learn a task embedding space 
from short programs and their I/O pairs 
that continuously parameterizes diverse program behaviors.
Then, we create a composed program dataset that provides 
intermediate supervision for learning a program composer, 
which efficiently learns to hierarchically compose short programs 
to form long and complex task-solving programs.
Experimental results on a string transformation domain demonstrate the effectiveness of our proposed framework.
Ablation studies provide detailed analysis on learned task embedding spaces 
as well as justify the proposed training schema that leverages a composed program dataset.

\newpage

% \bibliography{iclr2022_conference}

% \bibliographystyle{plain}
%\bibliographystyle{unsrt}
\renewcommand{\bibname}{References}
\bibliography{ref}
\bibliographystyle{iclr2023_conference}

\clearpage
\appendix

\section*{Appendix}

\begingroup
\hypersetup{
linkcolor=black,
pdfborder = {0 0 0},
}

\part{} % Start the appendix part
\parttoc % Insert the appendix TOC

% \listoffigures
% \listoftables
\endgroup

\section{Architecture}
\label{sec:arch}

\subsection{Task Encoder}
\label{sec:task_encoder}
The task encoder $e_\beta$ is a recurrent neural network 
which encodes an I/O task specification $\sigma$ to a task embedding. The encoded representation is then used by the program composer $h_\omega$
to synthesize programs. Specifically, the encoder is a gated recurrent unit (GRU) module with an input feature size of 256 and a hidden state size of 512. An embedding vector is calculated for the input string and separately for the output string. These embeddings are passed through a joint encoding module, which consists of a block of linear layers, dropout of 0.2, and rectified linear unit (ReLU) activation, followed by batch normalization.
 
\subsection{Program Decoder}
\label{sec:program_decoder}
First trained during the embedding space learning stage, the program decoder $p_\theta$ is a recurrent neural network 
which decodes a task embedding produced by the task encoder $e_\beta$ 
into a program $\hat{\rho}$. The model is trained such that the reconstructed program is as close as possible to the ground truth program whose task specification was the input to the task encoder. In HNPS, the program decoder takes as input the program embeddings from the program composer module and outputs program tokens. Under the hood, the decoder is a GRU module with an input feature size of 343 and a hidden state size of 256. The previously produced program tokens are encoded and concatenated with the program composer module output, passed through the GRU, and output logits are generated via two linear layers and the Tanh activation function. Finally, the program token is generated via a softmax over these logits.

\subsection{Program Composer}
\label{sec:program_composer}
The program composer $h_\omega$ is a recurrent neural network
which takes the output of the task interpreter as input and sequentially produce tasks embeddings. Specifically, at each step, $h_\omega$ takes in previously produced task embeddings, and last hidden state program decoder as input to
predict next task embedding. It consists of a GRU module with an input feature size of 768 and a hidden state size of 256. This is followed by a single block of a linear layer and leaky ReLU activation.

\section{Hyperparameters and Training Details}
\label{sec:training}
In the datasets ($D_\text{short}$, $D_\text{composed}$) on which we train our model, we use a training/validation/test split of 70\% / 15\% / 15\%. Our datasets contain 20 I/O string pairs for each program. During training, we randomly sample 10 of those 20 I/O pairs to use as the task specification.

We use the PyTorch framework for implementing HNPS.

\subsection{Learning a Program Embedding Space}
\label{sec:app_stage1}
In this stage, our model aims to generate a program that is as close as possible to the program whose behavior is described by the input task specification. To do this, we employ cross-entropy loss between the generated program tokens and the ground truth program tokens. We employ a training regime with teacher forcing behavior, where the previously predicted program tokens are replaced with the ground truth program tokens.

\subsection{Hierarchical Neural Program Synthesis}
\label{sec:app_hnps}
Here, in addition to the cross-entropy loss for individual program tokens, we employ a mean squared error loss for program embeddings. We use ground truth program embeddings for the program embedding loss during finetuning, but we let the embedding space drift. Even though ground truth program embeddings can be noisy, they still offer a valuable source of supervision for composing programs.

To determine when our synthesis module should stop outputting more program tokens, we initially tried to learn the prediction of a stop token. However, we observed that this was not robust. Instead, we take advantage of the fact that in the program synthesis domain, we can verify the execution of a program at any point. So, as we are composing programs together, if the addition of another subprogram causes a reduction in execution accuracy, then we terminate the program at this point - we found this heuristic performed well.

\subsection{Baseline Implementation Details}

Na\"{i}ve, Na\"{i}ve-short, and Na\"{i}ve-short-finetune are implemented with a GRU encoder-decoder model. The encoder is implemented with the same architecture and hyperparameters as the HNPS task encoder, and the decoder uses the same architecture and hyperparameters as the HNPS program decoder. 

The runtime threshold for the BUSTLE bottom-up search baseline is 500,000 expressions. Our results with this baseline are using the version of BUSTLE that leverages string heuristics to prioritize the sub-expression search. We experimented with variants that incorporated a neural network, but observed no increase in execution accuracy on long programs, so we just report results with the heuristic implementation.

\subsection{Hyperparameters}
Training runs use the Adam optimizer with an initial learning rate of 0.0001, and a learning rate scheduler decays the rate with a gamma of 0.95. A weight decay of 0.00001 is used. The model is optimized for 150 epochs of the dataset.

Hyperparameters used in our experiments are below:

\begin{itemize}
    \item Number of I/O String Pairs in Task Specification: 10
    \item Optimizer: Adam
    \item Learning Rate: 0.0001
    \item Learning Rate Scheduler Gamma: 0.95
    \item Weight Decay: 0.00001
    \item Training Epochs: 150
\end{itemize}

Task Encoder:
\begin{itemize}
    \item Model Type: GRU
    \item Model Type: GRU
    \item Input Size: 256
    \item Hidden State Size: 512
    \item Dropout: 0.2
    \item Activation Function: ReLU
    \item Number of Linear Layer Blocks: 1
    \item I/O Embedding Pooling Function: Mean
\end{itemize}

Program Decoder:
\begin{itemize}
    \item Model Type: GRU
    \item Input Size: 256
    \item Hidden State Size: 256
    \item Activation Function: Tanh
    \item Number of Linear Layers: 2
\end{itemize}

Program Composer:
\begin{itemize}
    \item Model Type: GRU
    \item Input Size: 768
    \item Hidden State Size: 256
    \item Activation Function: Leaky ReLU
    \item Number of Linear Layer Blocks: 1
\end{itemize}

HNPS Loss Coefficients:
\begin{itemize}
    \item $\lambda$: 1.0
    %\item $\lambda_2$: 1.0
\end{itemize}

\subsection{Syntax Checker}
\label{sec:syntax_checker}
Across all methods in our experiments, we utilize a syntax checker to make sure our synthesis module is outputting valid program tokens per the DSL. This aids in producing programs that can actually be compiled and executed. For example, if an open parentheses token is generated, the syntax checker will ensure that the expression in the parentheses is eventually closed with a close parentheses token. Another use case for the syntax checker is making sure regular expressions are formatted correctly during synthesis (\ie the regular expression function type must be one of the following: \texttt{Word}, \texttt{Num}, \texttt{Alphanum}, \texttt{Allcaps}, \texttt{Propcase}, \texttt{Lower}, \texttt{Digit}, \texttt{Char}.

\section{Program Dataset}
\label{sec:dataset_app}
Once we have trained the embedding space, we use the trained task encoder to fetch the embeddings of each of the short programs. These short programs are composed together for the dataset $D_\text{composed}$. We store the short program embeddings along with the programs that those short programs combine together to form, so that we can calculate our program embedding loss $\mathcal{L}^{\text{PE}}$.

\subsection{Program and Specification Generation Details}

In the generation process, we sample each program recursively using DSL grammar provided in Fig. \ref{fig:dsl}. Specifically, we would start with a program sketch with holes, Concat c( $e$ c) We then transform the hole, $e$, based on our grammar rules into ConstStr k( $s$ k), SubStr s( $q_1$ $q_2$ s), or into two expressions $e_1 e_2$. This process is repeated recursively until no holes are left in the program. The probabilities for each component we used in generation process are listed in Table.\ref{table:dsl_prob}. 

Given a generated program, we use a probabilistic sampler to create potential input strings. To decrease probability of exceptions in program execution, we use a simple set of heuristics for input string generation. For example, if a program includes Regex r( Digit 3 Start r) which calculate the location of the third number digit in input string, we would constrain the sampler to include at least 3 number digits in the input strings. 
\subsection{Dataset Statistics}
\label{sec:stats}
\begin{wraptable}[21]{r}{0.5\textwidth}
%\begin{table*}[h!]%[t]
    \centering
    \small
    \caption{\small 
    The probability of components being sampled to fill holes when generating each program. We use a recursive process for generation by filling holes in the current program with components specified in the DSL grammar rules. 
    }
    \resizebox{0.5\textwidth}{!}{
    \begin{tabular*}{\linewidth}{ll@{\extracolsep{\fill}}c}
    \toprule
    \textbf{Hole Type} & \textbf{Component} & \textbf{Probability} \\
    \midrule
    \multirow{3}{*}{Expression $e$}
    & ConstStr k( $s$ k)& 0.1 \\
    & SubStr s( $q_1$ $q_2$ s)& 0.35 \\
    & $e_1$ $e_2$ & 0.55 \\
    \midrule
    \multirow{2}{*}{Position $q$}
    & ConstPos p( $k$ p)& 0.1 \\
    & Regex r( $x$ $n$ $b$ r)& 0.35 \\
    \midrule
    Boundary $b$ & Start, End & 0.5 \\
    \midrule
    Regex $x$ & Word, Num, ..., Char & 0.0625 \\
    \midrule
    Position Index $k$ & -20, -19, ..., 20 & 0.0244 \\
    \midrule
    Position Index $k$ & -3, -2, ..., -3 & 0.143 \\
    \midrule
    Character $k$ & "SPACE", ".", ..."?" & 0.05 \\
    \bottomrule
    \end{tabular*}
	}
    \label{table:dsl_prob}
\end{wraptable}    
%\end{table*}
$D_\text{short}$ has 100,000 programs, which are created by randomly sampling one expression-long programs from our DSL. For each sampled program, we randomly generate 1,000 strings with 50-70 characters as potential input strings, and the first 20 input strings that can be executed without exceptions are stored in our dataset along with their corresponding output strings. If less than 20 input strings in the 1,000 generated are executed without exceptions for a program, then that program is discarded. For $D_\text{composed}$, we also generate 200,000 programs by composing sampled programs from $D_\text{short}$. In $D_\text{composed}$, 60,000 programs are composed with 2 short programs, 70,000 programs are composed with 3 short programs, and 70,000 programs are composed with 4 short programs. For each composed program in $D_\text{composed}$, the same method is used to generate I/O pairs. 

We generate a long program dataset, $D_\text{long}$, which consists of 30,000 programs, each containing 2-4 expressions by directly sampling from our DSL. $D_\text{long}$ contains 10,000 2 expression-long programs, 10,000 3 expression-long programs, and 10,000 4 expression-long programs. 

To demonstrate the distribution gap between the 3 datasets, in Table~\ref{table:token_frequency}, we show the frequency of important DSL tokens in each of our dataset. We calculate token frequency by counting total number of occurrences of each token divided by the amount of programs in the dataset. 
%\begin{wraptable}{R}{0.50\textwidth}
\begin{table*}[t]

    \centering
    \small
    \caption{\small Frequency of important DSL tokens in each of our datasets. We calculate token frequency by counting total number of occurrences of each token divided by the amount of programs in the dataset.
    }
    \scalebox{0.9}{
    \begin{tabular*}{\linewidth}{l@{\extracolsep{\fill}}ccc}
    \toprule
    token & $D_\text{short}$ & $D_\text{composed}$ & $D_\text{long}$\\
    \midrule
    Concat & 1.00 & 1.00 & 1.00 \\
	ConstStr & 0.06 & 0.23 & 0.11 \\
	SubStr & 0.94 & 2.82 & 2.89 \\
	Regex & 1.64 & 4.92 & 5.21 \\
	ConstPos & 0.23 & 0.72 & 0.57 \\
	Word & 0.09 & 0.25 & 0.32 \\
	Num & 0.04 & 0.11 & 0.34 \\
	Alphanum & 0.09 & 0.25 & 0.32 \\
	Allcaps & 0.08 & 0.25 & 0.33 \\
	Propcase & 0.09 & 0.26 & 0.33 \\
	Lower & 0.09 & 0.25 & 0.34 \\
	Digit & 0.09 & 0.26 & 0.32 \\
	Char & 0.06 & 0.21 & 0.31 \\
	All number tokens & 1.87 & 5.66 & 5.78 \\
	All char tokens & 1.08 & 3.31 & 2.71 \\
	Start & 0.82 & 2.46 & 2.60 \\
	End & 0.82 & 2.46 & 2.61 \\
    \bottomrule
    \end{tabular*}
	}
    \label{table:token_frequency}
%\end{wraptable}    
\end{table*}

\subsection{Program and I/O Pair Samples}

We include examples of programs from our $D_\text{composed}$ dataset and show 5 out of 20 I/O pairs we sampled for each program. 
\\
% \begin{figure}[t]
% \centering
%     \caption[]{
%         \small
%         Examples of programs and I/O pairs in $D_\text{composed}$ and programs generated by our HNPS model. For each program, we only show 5 out of the 20  I/O pairs sampled for each program
%         %A program consists of domain-dependent perceptions, actions, and control flows.
%         \label{fig:d_composed_ex}
%     }
\begin{mdframed}
{%\footnotesize
 \dslfontsize
 \begin{tabular*}{\linewidth}{ll}
    \toprule
    Input: & ZL\%@r\#ENc\#HMONNRZLEBHVZMYIDUYBZJWQPOQLVLXGLMREUJDXQI'aP@Q(dR(\#/E@7\#//\\ 
	Output: & @7ZL\%@r\#ENc\#HMONNRZLEBHVZMYIDUYBZJWQPOQLVLXGLMREUJDXQI'aP\\
	& @Q(dR(\#/E@7\#/\\ \\
 
	Input: & \#\#/@@@KA?/WYYESBKMODJKGDBBYWPVXFQINPRGVGYOHRQWDEQZIOCYVGPJh/\#\\ 
	Output: & @KA?/WYYESBKMODJKGDBBYWPVXFQINPRGVGYOHRQWDEQZIOCYVGPJh/KA?/W\\
	& YYESBKMODJKGDBBYWPVXFQINPRGVGYOHRQWDEQZIOCYVGPJh\\ \\
 
	Input: & ?!b:\{\#M@T@\$BRKOCJZARAAHMHPZWAKAUy:YSLDHLLVOPHFMPYTZCFr\#/X@///TLa\#\\ 
	Output: & @///TLaBRKOCJZARAAHMHPZWAKAUy:YSLDHLLVOPHFMPYTZCFr\#/X@//\\ \\
 
	Input: & @/EOJHGAWTNYWAETCPPFNTTECQVPFYSb@\#/\#@/\#EPGDIKWHJMVQNTVM3\\ 
	Output: & @/EOJHGAWTNYWAETCPPFNTTECQVPFYSb@\#/\#@\\ \\
 
	Input: & MG\#i/ZMTEMXUVXDKEMIDAFAUV8@QOOXGRGGFVQTSOZUAURSv@@7\#\%\#kX?/L[W'@z\{a\#/(\\ 
	Output: & @z\{aMG\#i/ZMTEMXUVXDKEMIDAFAUV8@QOOXGRGGFVQTSOZUAURSv@@7\#\\
	& \%\#kX?/L[W'@z\{a\#\\ \\
    \midrule
 \end{tabular*}
 \\
 \begin{tabular*}{\linewidth}{ll}
    Ground Truth Program: &
    Concat c( SubStr s( ConstPos p( 7 p) Regex r( "\#" 1 End r) s) \\
    & SubStr s( Regex r( "(" -2 Start r) Regex r( "." -2 End r) s) c) \\
    \bottomrule
 \end{tabular*}
}
\end{mdframed}

\begin{mdframed}
{%\footnotesize
 \dslfontsize
 \begin{tabular*}{\linewidth}{ll}
    \toprule
    Input: & G\#PZoez.Xo()\$D(AJW3Av\#q'@Xi?\#P..TlY(O(N\&XI\#G: Q.hQT\#dk)0Y(jSF\\ 
    Output: & .Xo()\$D(AJW3Av(N\&XI\#G: Q.hQT\#dk)0Y(j\\ \\

    Input: & xN\#iiu/HHfa4(!pEDz4\{q(,(nlw.\{JsGN\#SK;h./CjDt\}u.n0DgMuj\#KBR,H\#.(F;SR;R\\ 
    Output: & HHfa4(!pEDz4\{q(,(nlw.\{JsGN(nlw.\{JsGN\#SK;h./CjDt\}u.n0DgMuj\#KBR,H\#.(F;SR\\ \\
    
    Input: & ncL9.(Cbz3q[y9CXHkjjI84RA!sRX.T';\%G(kdk.@zb\#j(D\#X]\#.\&CHSwcdB([D\}4 1\\ 
    Output: & bz3q[y9CXHkjjI84RA!sRX.T';\%G(kdk.@zb\#j(D(D\#X]\#.\&CHSwcdB([D\}4\\ \\
    
    Input: & 78@aQ0(I(:\#[h]\#)8F3C.9(.5h9UXz\#\&L(2gZCR8d5iK33.Z.e[8\\ 
    Output: & I(:\#[h](.5h9UXz\#\&L(2gZCR8d5iK33.Z.e\\ \\
    
    Input: & :Jmq?T(R1hD.\#;,.(Q9X?q7Dm\}1Vq6e'0.J(\#\#C8]s]5.X/a0?mC9(5k\\ 
    Output: & R1hD.\#;,.(Q9X?q7Dm\}1Vq6e'0.J((\#\#C8]s]5.X/a0?mC9(\\ \\
    \midrule
 \end{tabular*}
 \\
 \begin{tabular*}{\linewidth}{ll}
    Ground Truth Program: &
    Concat c( SubStr s( Regex r( "@" -1 End r) Regex r( "\#" -1 End r) s) \\
    & SubStr s( Regex r( Allcaps 0 Start r) Regex r( "/" -1 End r) s) c) \\
    \bottomrule
 \end{tabular*}
}
\end{mdframed}

\begin{mdframed}
{%\footnotesize
 \dslfontsize
 \begin{tabular*}{\linewidth}{ll}
    \toprule
    Input: & !\&]dr]1rqFpih3CdBEEICIvPaNkuHa4\$0lX!\{IA\&eFkAUtbDBAs5GusY'\&\&\\ 
	Output: & !\&]dr]1rqFpih3CdBEEICIvPaNkuHa4\$0lX!\{IA\&eFkAUtbDBAs5Gus\\
	& \$]dr]1rqFpih3CdBEEICIvPaNkuHa4\$0lX!\{IA\&eFkAUtbDBAs5GusY'\\ \\
 
	Input: & a]\&]\&EqZWNGObvBcuKmtUvUzEtupBTYdHjfyMOj3R!f\&wkhWgLeLk5!tShuzGfo4\&\\ 
	Output: & !f\&wkhWgLeLk5!tShuzGf\$]\&]\&EqZWNGObvBcuKmtUvUzEtupBTYdHjf\\
	& yMOj3R!f\\ \\
 
	Input: & evFjtgoxhRAUsAoeAbNK4\&\$m]SfjBenvmUL9]!.MQBWzpFEbIcktLQSgasdlHc\#\&\&Eg\&!e\\ 
	Output: & !.MQBWzpFEbIcktLQSgasdlHc\#\&\&E\$]SfjBenvmUL9]!.MQBWzpFEbIc\\
	& ktLQSgasdlHc\#\&\\ \\
 
	Input: & f\&]LP!rSAlV\&INLbdV!\&.soP\%D@YDxCzkUaC63IgUV7sLsgxzcgnFsbWCz7!D2H DqA\&]\\ 
	Output: & !rSAlV\&INLbdV!\&.soP\%D@YDxCzkUaC63IgUV7sLsgxzcgnFsbWCz7!D2\\
	& H Dq\$]LP!rSAlV\&INLbdV!\\ \\
	
	Input: & Z](\&SCh2!\&(SWbZZjJOD1PC]FlgEAwCAXNZsLbdmJbzFLwd\$V!\&?4n\&qVVQ\\ 
	Output: & !\&(SWbZZjJOD1PC]FlgEAwCAXNZsLbdmJbzFLwd\$V!\&?4n\&qVV\$](\&\\
	& SCh2!\&(SWbZZjJOD1PC]FlgEAwCAXNZsLbdmJbzFLwd\$V!\\ \\
 
    \midrule
 \end{tabular*}
 \\
 \begin{tabular*}{\linewidth}{ll}
    Ground Truth Program: &
    Concat c( SubStr s( Regex r( "!" 0 Start r) Regex r( Word -1 End r) s) \\
    & ConstStr k( "\$" k) \\
    & SubStr s( Regex r( "]" 0 End r) Regex r( "\&" 2 End r) s) c) \\
    \bottomrule
 \end{tabular*}
}
\end{mdframed}

\begin{mdframed}
{%\footnotesize
 \dslfontsize
 \begin{tabular*}{\linewidth}{ll}
    \toprule
    Input: & ;(qx\{Y,,,FkAUtb\}))D))DCAt1W,(2,S;BAspoCdBEEICIvPaNkuHajGdqFpi4;:F\\ 
	Output: & (qx\{Y,,))DCAFkAUtb\}))D))DCAt1W,(2,S\&\\ \\
 
	Input: & VI,r\#;zGfo6,q);/X(3,,roqpvaDzj0,R;Z')(\&yK;ZDEq3)OSus\$xGc)AqskKS7BX\{lh\\ 
	Output: & (3)OSus\$xGc)AqskKS7BzGfo6,q)\&\\ \\
 
	Input: & )S),);NpQFmcNGTMMQB4bbXkWitgy0,;(S,o;,)LmjUGnwzcm7,(IPpqrv10\\ 
	Output: & (S);NpQFmcNGTMMQB4bbXkWitgy0,;(S,o;,)LmjUGnwzcQFmcNGTMMQB4bb\\
	& XkWitgy0,\&\\ \\
 
	Input: & HW(fMhJyyV0)T)ex,),iLiRno'E;;,),)yZXAIbscuV6(vUHgjyB7,;o\\ 
	Output: & (fMhJyyV0)T)ex,),iLiRno'E;;),)yZXAIbscuMhJyyV0)T)ex,),iLiRno\\
	& 'E;\&\\ \\
 
	Input: & )zRxForqNOFp7FL,4))2;ksgGpWkZRRdGMvjxaT2((,,;1;4)jYeft9,FYrrzBY7,b\\ 
	Output: & ((,)2;ksgGpWkZRRdGMvjxaT2((,,;1;4)jYefForqNOFp7FL,4))2;ksgGp\\
	& WkZRRdGMvjxaT2((,,\&\\ \\
    \midrule
 \end{tabular*}
 \\
 \begin{tabular*}{\linewidth}{ll}
    Ground Truth Program: &
    Concat c( SubStr s( Regex r( "(" 0 Start r) Regex r( "," -3 Start r) s) \\
    & SubStr s( Regex r( ")" -2 Start r) Regex r( Word -2 End r) s) \\
    & SubStr s( Regex r( Char 3 End r) Regex r( ";" 1 End r) s) \\
    & ConstStr k( "\&" k) c) \\
    \bottomrule
 \end{tabular*}
}
\end{mdframed}

\begin{mdframed}
{%\footnotesize
 \dslfontsize
 \begin{tabular*}{\linewidth}{ll}
    \toprule
    Input: & 2q@?QK6IAd\}t2?))\}!\}!l?AV\}UP,Hnriz1OoH?!CsyjyICX!!!Esi\$80)DU5i\}\}\\ 
	Output: & @?QK6IAd\}t2?))\}!\}!l?AV\}UP,Hnriz1OoH?!CsyjyICX))\}!\}!l?A\\
	& V\}Ut2?))\}!\}!l?AV?!\\ \\
 
	Input: & !WHj!?XUqQT \}\}yOI)x1[SWU'!KoVWd.RoutoK\}!?\}?Rocescz7!lWBxhk\$OXu)\}a))\\ 
	Output: & Hj!?XUqQT \}\}yOI)x1[SWU'!KoVWd.RoutoK\})x1[SWU'!KoVyOI)x1[S\\
	& WU'!KoVWd.RoutoK?Rocescz7!lW\\ \\
 
	Input: & Xcxu6EQ1!Bx(\}\}1KMx?!\}8U@\}?!!?M)XpoeCWokeC!Lbiai.EQ7un\}BH!\})zy\{QA(\\ 
	Output: & xu6EQ1!Bx(\}\}1KMx?!\}8U@\}?!)XpoeCWokeC!Lbiai.E1KMx?!\}8U@?\\
	& M)XpoeC\\ \\
 
	Input: & EAb?u6!)DK]!Cp\#?mFo\$IR\&cC/\})Y!3Imegp3SW@CV(!\}n\}?MseO\}Jbdt3W!Tr7\}BKp\\ 
	Output: & b?u6!)DK]!Cp\#?mFo\$IR\&cC/\})Y!3Imegp3SW@CV()DK]!Cp\#?mFo\$\\
	& IR\&cC/\})Y!3Imegp3SDK]!Cp\#?mFo\$IR\&cC/\})Y!3Imegp3SW@CV(!\}n?MseO\}Jbdt3W!\\ \\
 
	Input: & CR@Vo,!?!))!NL'PBoSZ1\}!ToKOOoja\$TphjC!\}BjhfwkJJO(?q\}?\}DrpRtNN2\}\\ 
	Output: & @Vo,!?!))!NL'PBoSZ1\}))!NL'PBoSZ1\}!ToKONL'PBoSZ1\}!ToKOOoja\\
	& \$TphjC!\}BjhfwkJJO(?q?\}\\ \\
    \midrule
 \end{tabular*}
 \\
 \begin{tabular*}{\linewidth}{ll}
    Ground Truth Program: &
    Concat c( SubStr s( ConstPos p( 2 p) Regex r( "!" 3 Start r) s) \\
    & SubStr s( Regex r( ")" 0 Start r) Regex r( Allcaps -3 End r) s) \\
    & SubStr s( Regex r( Alphanum 2 Start r) Regex r( "\}" -3 End r) s) \\
    & SubStr s( Regex r( "?" -1 Start r) Regex r( Propcase -2 Start r) s) c) \\
    \bottomrule
 \end{tabular*}
}
\end{mdframed}
% \end{figure}

\label{sec:sample}

\section{Extended Results}

\subsection{Generated Programs}
We show examples of programs generated by our HNPS model. Each program is generated given task specifications randomly picked from $D_\text{composed}$. We show a comparison between each generated program with the ground truth program given by the datsaset.
\\
\begin{mdframed}
{%\footnotesize
 \dslfontsize
 \begin{tabular*}{\linewidth}{ll}
    Ground Truth Program: &
    Concat c( SubStr s( Regex r( Alphanum -3 Start r) Regex r( "(" -2 Start r) s) \\
    & SubStr s( Regex r( Alphanum -1 End r) Regex r( "." -1 Start r) s) c) \\
    \midrule
    HNPS Generated Program: &
    Concat c( SubStr s( Regex r( Alphanum -3 Start r) Regex r( "(" -2 End r) s) \\
    & SubStr s( Regex r( Alphanum -1 End r) Regex r( "." -1 Start r) s) c) \\
 \end{tabular*}
}
\end{mdframed}

\begin{mdframed}
{%\footnotesize
 \dslfontsize
 \begin{tabular*}{\linewidth}{ll}
    Ground Truth Program: &
    Concat c( SubStr s( Regex r( "!" 0 Start r) Regex r( "\$" -1 End r) s) \\
    & SubStr s( ConstPos p( 12 p) Regex r( "." -1 End r) s) c) \\
    \midrule
    HNPS Generated Program: &
    Concat c( SubStr s( Regex r( "!" 0 End r) Regex r( "." -1 End r) s) \\
    & SubStr s( ConstPos p( 12 p) Regex r( "\$" -1 End r) c) \\
 \end{tabular*}
}
\end{mdframed}

\begin{mdframed}
{%\footnotesize
 \dslfontsize
 \begin{tabular*}{\linewidth}{ll}
    Ground Truth Program: &
    Concat c( SubStr s( Regex r( Propcase -3 End r) Regex r( "[" -2 Start r) s) \\
    & SubStr s( Regex r( "." 1 Start r) Regex r( Propcase 1 Start r) s) \\
    & SubStr s( Regex r( "@" -3 End r) ConstPos p( -8 p)) c) \\
    \midrule
    HNPS Generated Program: &
    Concat c( SubStr s( Regex r( Propcase -3 End r) Regex r( "[" -2 End r) s) \\
    & SubStr s( Regex r( "." 1 End r) Regex r( Propcase 1 Start r) s) \\
    & SubStr s( Regex r( "@" -3 Start r) ConstPos p( -8 p) c) \\
 \end{tabular*}
}
\end{mdframed}

\begin{mdframed}
{%\footnotesize
 \dslfontsize
 \begin{tabular*}{\linewidth}{ll}
    Ground Truth Program: &
    Concat c( SubStr s( Regex r( "SPACE" -1 End r) Regex r( "\{" -2 End r) s) \\
    & SubStr s( Regex r( "?" 2 End r) Regex r( Lower 2 End r) s) \\
    & SubStr s( Regex r( Allcaps 3 Start r) Regex r( "," -1 Start r) s) c) \\
    \midrule
    HNPS Generated Program: &
    Concat c( SubStr s( Regex r( "SPACE" -1 Start r) Regex r( "\{" -2 End r) s) \\
    & SubStr s( Regex r( "?" -2 Start r) Regex r( Lower -1 End r) s) \\
    & SubStr s( Regex r( Allcaps 3 Start r) Regex r( "," -1 Start r) s) c) \\
 \end{tabular*}
}
\end{mdframed}

\begin{mdframed}
{%\footnotesize
 \dslfontsize
 \begin{tabular*}{\linewidth}{ll}
    Ground Truth Program: &
    Concat c( SubStr s( ConstPos p( 14 p) Regex r( "\$" 0 End r) s) \\
    & SubStr s( Regex r( "'" -2 Start r) Regex r( Digit -2 Start r) s) \\
    & SubStr s( Regex r( Digit 0 End r) Regex r( Word 3 End r) s) \\
    & SubStr s( Regex r( "!" 2 Start r) Regex r( Char -3 Start r) s c) \\
    \midrule
    HNPS Generated Program: &
    Concat c( SubStr s( ConstPos p( 14 p) Regex r( Char -1 End r) s) \\
    & SubStr s( Regex r( "'" 2 End r) Regex r( Digit -2 Start r) s) \\
    & SubStr s( Regex r( Char 3 End r) Regex r( Word 3 End r) s) \\
    & SubStr s( Regex r( "!" 2 Start r) ConstPos p( -4 p) c) \\
 \end{tabular*}
}
\end{mdframed}

\begin{mdframed}
{%\footnotesize
 \dslfontsize
 \begin{tabular*}{\linewidth}{ll}
    Concat c( ConstStr k( "SPACE" k) \\
    & SubStr s( Regex r( "." 2 End r) Regex r( "?" 2 End r) s) \\
    & SubStr s( Regex r( "@" -2 End r) Regex r( Alphanum 3 End r) s) \\
    & SubStr s( Regex r( ")" 1 Start r) Regex r( Num -1 Start r) s) s) c) \\
    \midrule
    HNPS Generated Program: &
    Concat c( ConstStr k( "SPACE" k) \\
    & SubStr s( Regex r( "." 2 End r) Regex r( "?" 2 End r) s) \\
    & SubStr s( Regex r( "@" -2 End r) Regex r( Alphanum 3 End r) s) \\
    & SubStr s( Regex r( ")" 1 End r) Regex r( Num -1 Start r) s) c) \\
 \end{tabular*}
}
\end{mdframed}

\subsection{Evaluation on $D_\text{composed}$}
\label{sec:d_composed_eval}
 We additionally evaluate execution accuracy of each method on our $D_{composed}$ dataset. In Table~\ref{table:d_composed_token_num}, we group programs by their length in terms of the number of tokens. In Table~\ref{table:d_composed_program_num}, we separate programs by the number of short programs used to composed them. From the results, we can see HNPS \& HNPS-PT are able to outperform H-Na\"{i}ve-PT. The gap in performance between these and H-Na\"{i}ve-PT increases as programs get longer. For long programs composed of 2 short programs, H-Na\"{i}ve-PT reaches 93\% of HNPS's performance, but its execution accuracy is only 60\% of that of HNPS for long programs composed of 4 short programs. This highlights the importance of the task embedding space in program synthesis with hierarchical architectures. We also observe that non-hierarchical baselines (Na\"{i}ve-*) have significantly lower performance than hierarchical methods on both training and test set programs. On programs composed of 4 short programs, none of the non-hierarchical baselines are able to achieve an execution accuracy higher than 2\%, even on the training set, where all  hierarchical variants reach more than 10\% execution accuracy. This supports our assertion that without the hierarchical architecture, the model's ability to fit to a distribution of long programs is fundamentally limited. 

In Table~\ref{table:generalization_gap}, we also present the difference between training and test set performance, or the ``generalization gap,'' on $D_\text{composed}$. HNPS has a smaller generalization gap than both HNPS-PT and H-Na\"{i}ve-PT. This difference is especially noticeable when comparing HNPS and HNPS-PT on programs longer than 40 tokens, where HNPS-PT performs worse on the test set despite obtaining a better execution accuracy on the training set. This implies that the program embedding loss helps prevent overfitting in hierarchical models. 
\begin{table}[t]
%\begin{wraptable}{r}{0.55\textwidth}
\begin{subtable}
    \centering
    \small
    \caption{\small 
    The execution accuracy of each method on the training set and test set of $D_\text{composed}$. The programs are separated into groups by their length in terms of number of tokens.
    }
    \scalebox{1.0}{
    \begin{tabular*}{\linewidth}{l@{\extracolsep{\fill}}cccccccc}
        \toprule
       \multirow{2}{*}{\textbf{Method}} 
       &\multicolumn{4}{c}{\textbf{Train}}&\multicolumn{4}{c}{\textbf{Test}}\\ 
        \cmidrule(lr){2-5}\cmidrule(lr){6-9} 
        &\textbf{10-25} &\textbf{25-40} &\textbf{40-55} &\textbf{55-70}
        &\textbf{10-25} &\textbf{25-40} &\textbf{40-55} &\textbf{55-70}
        \\
        % &Program &Execution &Program &Execution \\

        \midrule
        Na\"{i}ve &54.55\% &21.73\% &2.22\% &0.05\% &50.08\% &17.8\% &1.64\% &0.01\% \\
        Na\"{i}ve-short &0.00\% &0.00\% &0.00\% &0.00\% &0.00\% &0.00\% &0.00\% &0.00\% \\
        Na\"{i}ve-short-finetune &56.05\% &28.37\% &6.10\% &0.34\% &53.82\% &24.21\% &3.87\% &0.18\% \\
    
        \midrule
        H-Na\"{i}ve-PT &77.82\% &51.62\% &26.62\% &8.39\% &69.54\% &37.48\% &11.35\% &1.04\% \\

        \midrule
        HNPS-PT &\textbf{82.50\%} &\textbf{59.84\%} &\textbf{33.09\%} &\textbf{11.72\%} &\textbf{74.05}\% &\textbf{45.04\%} &15.59\% &1.80\% \\ 
        HNPS (ours-full) &80.74\% &56.75\% &30.00\% &9.66\% &73.36\% &43.88\% &\textbf{15.70\%} &\textbf{2.30\%} \\
        \bottomrule
    \end{tabular*}
	}
    \label{table:d_composed_token_num}
%\end{wraptable}    
\end{subtable}

%\begin{wraptable}{r}{0.55\textwidth}
\begin{subtable}
    \centering
    \small
    \caption{\small 
    The execution accuracy of each method on the training set and test set of $D_\text{composed}$. The programs are separated into groups by the amount of subprograms used to compose them.
        }
    \scalebox{1.0}{
    \begin{tabular*}{\linewidth}{l@{\extracolsep{\fill}}cccccc}
        \toprule
       \multirow{2}{*}{
       \textbf{Method}} &\multicolumn{3}{c}{\textbf{Train}}&\multicolumn{3}{c}{\textbf{Test}}\\ 
        \cmidrule(lr){2-4}\cmidrule(lr){5-7} 
        &\textbf{2} &\textbf{3} &\textbf{4} &
        \textbf{2} &\textbf{3} &\textbf{4}  \\

        % &Program &Execution &Program &Execution \\

        \midrule
        Na\"{i}ve &27.30\% &5.83\% &0.60\% &22.99\% &4.27\% &0.51\% \\
        Na\"{i}ve-short &0.02\% &0.00\% &0.00\% &0.00\% &0.00\% &0.00\% \\
        Na\"{i}ve-short-finetune &34.66\% &10.06\% &1.08\% &30.10\% &7.16\% &0.53\% \\
    
        \midrule
        H-Na\"{i}ve-PT &56.08\% &32.12\% &13.19\% &43.15\% &16.81\% &3.02\%\\

        \midrule
        HNPS-PT &\textbf{64.10\%} &\textbf{39.95\%} &\textbf{16.25\%} &\textbf{50.80\%} &21.85\% &4.26\% \\
        HNPS (ours-full) &61.09\% &36.27\% &14.55\% &48.96\% &\textbf{22.02\%} &\textbf{5.06\%} \\
        \bottomrule
    \end{tabular*}
	}
    \label{table:d_composed_program_num}
%\end{wraptable}    
\end{subtable}

%\begin{wraptable}{r}{0.55\textwidth}
\begin{subtable}
    \centering
    \small
    \caption{\small 
        We show the ``generalization gap'' of each method by calculating the the difference between training and test set performance.
        }
    \scalebox{1.0}{
    \begin{tabular*}{\linewidth}{l@{\extracolsep{\fill}}ccccccc}
        \toprule
       \multirow{2}{*}{
       \textbf{Method}} &\multicolumn{4}{c}{\textbf{By number of tokens}}&\multicolumn{3}{c}{\textbf{By number of short programs}}\\ 
        \cmidrule(lr){2-5}\cmidrule(lr){6-8} 
        &\textbf{10-25} &\textbf{25-40} &\textbf{40-55} &\textbf{55-70} &
        \textbf{2} &\textbf{3} &\textbf{4}  \\

        % &Program &Execution &Program &Execution \\

        \midrule
        Na\"{i}ve &-4.48\% &-3.94\% &-0.58\% &-0.04\% &-4.32\% &-1.56\% &-0.09\% \\
        Na\"{i}ve-short &0.00\% &0.00\% &0.00\% &0.00\% &0.00\% &0.00\% &0.00\% \\
        Na\"{i}ve-short-finetune &-2.23\% &-4.16\% &-2.22\% &-0.16\% &-4.57\% &-2.9\% &-0.55\% \\
    
        \midrule
        H-Na\"{i}ve-PT &-8.28\% &-14.14\% &-15.26 \% &\textbf{-7.35\%} &-12.93\% &-15.31\% &10.17\%\\

        \midrule
        HNPS-PT &-8.46\% &-14.8\% &-17.5\% &-9.92\% &-13.3\% &-18.1\% &-11.99\\
        HNPS (ours-full) &\textbf{-7.38\%} &\textbf{-12.87\%} &\textbf{-14.3\%} &-7.36\% &\textbf{-12.13\%} &\textbf{-14.25\%} &\textbf{-9.49\%} \\
        \bottomrule
    \end{tabular*}
	}
    \label{table:generalization_gap}
%\end{wraptable}    
\end{subtable}
\end{table}

\end{document}